
\input phyzzx
\hoffset=0.2truein
\voffset=0.1truein
\hsize=6truein
\def\TITLEPAGE{\frontpagetrue}
\def\CfA{
       \centerline{Harvard-Smithsonian Center for Astrophysics}
\centerline{60 Garden Street, Cambridge, MA 02138}
}
\def\TITLE#1{\vskip .5in \centerline{\fourteenpoint #1}}
\def\AUTHOR#1{\vskip .2in \centerline{#1}}

\def\ABSTRACT#1{\vskip .2in \vfil \centerline{\twelvepoint
\bf Abstract}
   #1 \vfil}
\def\ENDTITLEPAGE{\vfil\eject\pageno=1}
\hfuzz=5pt
\tolerance=10000
\TITLEPAGE
\TITLE{ELECTROWEAK STRINGS PRODUCE BARYONS
}

\AUTHOR{Manuel Barriola}

\CfA

{\baselineskip=14pt\ABSTRACT{
We propose a new mechanism to generate baryons.
Electroweak strings produced at the
electroweak phase transition will introduce anomalous currents
through interactions of the strings with the background
electromagnetic field and through changes in the helicity of the string
network. These anomalous currents  will produce
fluctuations in the baryon and lepton number.
We show that in the two-Higgs model corrections to the action coming
from the effective potential introduce a bias in the baryon
asymmetry that leads to a net baryon number production.
}}
\ENDTITLEPAGE
\eject

\REF\sa{A. D. Sakharov JETP Lett. 5:24 (1967).}
\REF\ck{A.G. Cohen, D.B. Kaplan and A.E. Nelson,
Annu. Rev. Nucl. Part. Sci {\bf 43} 27 (1993).}
\REF\nt{N. Turok, in `Perspectives on Higgs Physics', ed. G. Kane
(World Scientific, Singapore, 1992).}
\REF\do{A.D. Dolgov UM-AC.93-91 preprint (1993).}
\REF\nm{N. S. Manton Phys. Rev. D {\bf 28}, 2019 (1983): F. R.
  Klinkhammer and N. S. Manton, Phys Rev. D {\bf 30}, 2212 (1984).}
\REF\bd{R. H. Brandenberger and A.C. Davis,
 Phys. Lett. {\bf B308}, 79 (1993).}
\REF\cc{See \ck and references therein. }
\REF\bd{R. Brandenberger and A. C. Davis, Phys. Lett. {\bf B308}, 79
 (1993).}
\REF\br{R. Brandenberger, A. C. Davis and M. Trodden, BROWN-HET-935
preprint (1994).}
\REF\t{T. Vachaspati, Phys. Rev. Lett. {\bf 68}, 1977 (1992).
For earlier work see Y. Nambu, Nucl. Phys. {\bf B130}, 505 (1977);
M.B. Einhorn and A. Savit, Phys. Lett. {\bf 77B}, (1978) 295;
N.S. Manton, Phys. Rev. {\bf D28}, (1983) 2018.}
\REF\ej{M. A. Earnshaw and M. James Phys. Rev. {\bf d48}, 5818 (1993).}
\REF\lp{ L. Perivolaropoulos Phys. Lett. {\bf B316}, 528 (1993).}
\REF\vw{T. Vachaspati and R. Watkins, Phys. Lett. {\bf B318}, 163 (1993);
T.D. Lee, Phys. Rep. {\bf 23C}, 254 (1976).}
\REF\v{G. Dvali and G. Senjanovic, Phys. Rev. Lett. {\bf 71}, 2376 (1993).}
\REF\s{G. Dvali and G. Senjanovic  IC/94 preprint (1994).}
\REF\hg{ C.J. Hogan, Phys. Rev. Lett. {\bf 51} (1983) 1488;
 M.S. Turner and L.W. Widrow, Phys. Rev. {\bf D37} (1988);
 W. Garretson, G.B. Field and S.M. Carroll, Phys. Rev. {\bf D46}
 (1992) 5346.}
\REF\tv{T. Vachaspati, Phys. Lett. {\bf B265}, 258 (1991).}
\REF\eo{K. Enqvist and P. Olesen NORDITA-94/6 preprint (1994).}
\REF\sav{G.K. Savvidy, Phys. Lett. {\bf B71} (1977) 133.}
\REF\gf{I want to thank George Field for pointing this out when
I showed him the calculation.}
\REF\vf{T. Vachaspati and G. Field, TU/94 preprint (1994).}
\REF\tz{N. Turok and J. Zadrozny, Phys. Rev. Lett. {\bf 65}, 2331, (1990).}
\REF\dt{M. Dine and S. Thomas SCIPP 94/01 preprint (1994).}
\REF\mt{See D. Mitchel and N. Turok, Nucl. Phys. B {\bf 294}, 1138 (1987)
and references therein.}

One implication of the standard hot big bang cosmology is
that from the relics of the early universe we should in principle
expect the universe to contain the same abundance of baryons and
antibaryons. However there is compelling empirical evidence that suggests
the universe is made out of matter with a relatively small amount of
antimatter.
{}From the present asymmetry we can extrapolate that at the energy
scale of the electroweak phase transition there was one part in $10^8$
more matter than antimatter in the universe.
The goal of baryogenesis
is to explain this asymmetry between the amount of
matter and antimatter.

Sakharov\refmark\sa  realized that there are three necessary
conditions for any baryogenesis mechanism. First we need the
existence of interactions that violate baryon number. The second condition
is that the theory in consideration biases processes that violate
baryon number towards the production of a net baryon number. This requires
interactions that break the symmetries of
charge conjugation ($C$) and the product of charge
conjugation and parity ($CP$). Finally the
interactions have to be in the presence of processes that
are out of thermal equilibrium.

Recently it has been suggested that baryogenesis takes place at the
electroweak phase transition (for a review see Refs.\ck ,\nt and \do ).
The three main ingredients of this picture are:

First, in the electroweak model we know that in the field configuration
space the sphaleron\refmark\nm is the saddle point configuration connecting
two different minima. These minima correspond to vacuum configurations
with different baryon numbers. Transitions among the  minima will change
the baryon number.

Second, in order to produce a net baryon number asymmetry
the electroweak interactions must violate $C$ and $CP$. Extensions of the
Minimal Standard Model (MSM) are required
because the MSM violates $C$ and $CP$
in amounts too small to explain the baryon to antibaryon asymmetry.

Finally if the phase
transition is first order, the bubbles produced will nucleate and expand.
Near the expanding bubbles we shall have processes that depart from thermal
equilibrium.

Different calculations of the characteristics of the phase
transition\refmark\cc seem to show that the transition is first order
for small Higgs masses. The strength of the transition decreases
as the Higgs mass increases to the point at which the transition is
second order. In particular for the minimal standard electroweak
model, the present lower bound of the Higgs mass (60 Gev) suggests that
the phase transition is weakly first order.
It also seems likely that in this range for the Higgs mass, the
amount of anomalous baryon number violation after the phase transition will
wash out any baryon number that was created during the phase transition.
Finally we point out that if the transition is second order
baryogenesis will not take place since
the transition will be in thermal equilibrium.

Recently Brandenberger and Davis\refmark\bd
have proposed a new mechanism that uses
strings produced at the electroweak phase transition.
At the core of the
strings we have trapped false vacuum in which transitions that change baryon
number can occur through
configurations that connect two vacua with different baryon number.
The collapsing strings will be a source for the interactions that
violate $CP$ and of processes that are
out of thermal equilibrium. In order for
their  mechanism to work the thickness of the string should be bigger
than the magnetic screening length at high temperatures,
so that the transitions that change baryon number can take place.
This requires that the string core be thick
so the Higgs mass is subsequently small. This implies that the
phase transition will be more likely first order.
Ref.\br ~ also studies strings but these are produced at a symmetry
breaking whose energy is slightly higher
than the electroweak scale. The electroweak symmetry is
unbroken at the core of the cosmic strings and transitions that change
baryon number proceed as before.

In this paper we would like to review briefly
electroweak strings (in particular Z-strings\refmark\t )
and different mechanisms that imply magnetic fields at the time of the
electroweak phase transition.
We shall show how anomalous currents occur in the electroweak model.
We shall examine the interaction
of the magnetic Z-flux along the core of the string with the background
magnetic field. This interaction will introduce
baryon and lepton anomalous currents. Finally we shall study their
implications as a possible mechanism for baryogenesis.

Ref.\t ~ shows that in the standard electroweak model there are
solutions that look like Nielsen-Olesen strings with magnetic
Z-flux along their cores. Nambu\refmark\t has also shown that these
strings can connect monopole-antimonopole pairs.
The strings are metastable for some range in parameter space
even though the standard electroweak theory is topologically trivial.
However they are not stable for the physical value of the Weinberg angle,
$\theta _w $.
We should expect that Z-string solutions survive for
extensions of the standard electroweak model. For example Refs.\ej and \lp ~
have shown that this is the case in the two-Higgs model extension.
Ref.\ej ~ has also
studied the stability of this solution to small perturbations. They find
that for realistic values of the Higgs mass and the Weinberg angle the
string is unstable.

Dvali and Senjanovic\refmark\v have also shown that if we
consider the electroweak model with two Higgs doublets
and we add an extra global $U(1)_{gl}$ symmetry to the theory,
then there are topologically stable string solutions that also
carry magnetic Z-flux along their core.
The global $U(1)$ symmetry is obtained when we neglect the $CP$ violating
terms of the two-Higgs potential.
These terms have the generic form
$V_{CP}= \lambda (\Phi_1^\dagger \Phi_2 - \eta_1 \eta_2 e^{i\theta})$,
where $\theta$ is the $CP$ violating phase.
Therefore the only $CP$ violation with these strings will come from
the $KM$ matrix and this has been shown to be too small to explain
the baryon asymmetry.
However these strings will be metastable for sufficiently small
$\lambda$. This possibility is presently being investigated.
Ref.\s  ~ shows that
the minimal supergravity extension of the standard model has these
topological solutions if the hidden sector has an exact R-symmetry.
These strings have
different characteristics than `ordinary' embedded
Z-strings produced at the time of the phase transition. We shall not
consider them in this paper.

In what follows we will consider the embedded Z-strings
solutions of Ref.\t ~. We shall assume extensions of the minimal
standard model in which these strings are either stable or metastable.
These strings will be stabilized by the
nature of this baryogenesis mechanism. As we shall see, at the core of the
strings massless baryons will
be produced while outside they would be massive.
Ref.\vw ~ has shown that bound states can improve the stability of
nontopological solitons. However we should point out that no model
has yet been found in which these strings are metastabe for realistic
values of the Higgs mass and the Weinberg angle.

Since the
electroweak model has monopole solutions we expect that when
our string network is produced we will have monopole-antimonopole
($M-\bar M$) pairs
connected by strings and small string loops.
The $M-\bar M$ and strings will interact with the background plasma and
the monopoles will contract with some speed $v_c$.
If $l$ is the typical length of the strings in the network then
after the time interval $\delta t_d = l/v_c$, the strings will decay.
The strings will also have some transverse velocity with respect to
the background that we will denote by $v_t$.

There are several mechanism that will generate primordial magnetic
fields in the early universe. Some of them relay on physical effects during an
inflationary phase transition\refmark\hg .
It has also been suggested\refmark\tv
that in the electroweak phase transition the
fields have to be uncorrelated at distances larger than
the initial correlation length($\xi$).
Therefore the gradients of the
Higgs fields can not be compensated by the gauge fields for scales
bigger than $\xi$.
These gradients imply that we will have electromagnetic fields.
Recently Ref.\eo ~ has used the Savvidy vacuum\refmark\sav
to generate magnetic fields at the scale of grand unification.
They propose that magnetic field fluctuations
at the GUT scales produce a phase transition to a new ground
state with a non zero magnetic field. All these mechanisms imply strong
magnetic fields at the time of the electroweak phase transition.

Now that we have reviewed electroweak strings and the existence of
magnetic fields at the time of the phase transition, we shall show
how the anomalous currents form in this model.
In the electroweak model all the currents that couple with the gauge fields
are free of anomalies so that the theory is renormalizable.
However the baryon and lepton currents have anomalies of the form
$$
\partial_\mu J^{\mu}_B = \partial_{\mu} J^\mu _L =
N_f \left( {g^2 \over {32 \pi ^2 }} W^{\mu \nu}_a \tilde W^a _{\mu \nu} -
{g'^2 \over {32 \pi ^2 }} Y_{\mu\nu} \tilde Y^{\mu\nu} \right).
\eqn\anom
$$
Where $N_f$ is the number of families, $W^{\mu \nu}_a$ $a=1,2,3$
and $Y_{\mu \nu}$ are the
$SU(2)$ and $U_{Y}(1)$ field strengths respectively, with the tilde referring
to the usual definition of the dual of the field strength. $g$ and $g'$
are the associated gauge couplings. Note that there is not an anomaly for
the difference of the baryon to the lepton currents.

If we integrate both sides of Eq.\anom over a volume V and assume that
the currents vanish at the surface of V we obtain that the baryon number
$B=\int d^3 x J^0 _B$ changes in some time interval by,
$$
\Delta B = {N_f \over {32 \pi ^2}} \int dt \int d^3 x \left(
         g^2 W^{\mu \nu}_a \tilde W^a _{\mu \nu} -
                g'^2 Y_{\mu\nu} \tilde Y^{\mu\nu} \right).
\eqn\intg
$$
We see that if the r.h.s. of Eq.\intg ~ is non zero we shall have a change
in the baryon number.
Another way to interpret this result is realizing that the integrand
can be
expressed as a total divergence so if we neglect surface effects and
define the Chern-Simons numbers $N_{cs}$ and $n_{cs}$
$$
\eqalign{
N_{cs} &= {g^2 \over{32 \pi ^2}} \int d^3 x \epsilon ^{ijk}
  \left ( W^a _{ij} W^a _k - {1 \over3} g \epsilon _{abc}
              W^a _i W^b _j W^c _k \right ) \cr
n_{cs} &= {g'^2 \over {32 \pi ^2}} \int d^3 x
             \epsilon ^{ijk} Y_{ij} Y_k }
\eqn\simo
$$
we can perform the time integral to obtain
$$
\Delta B = N_f (\Delta N_{cs} - \Delta n_{cs}).
\eqn\cher
$$
Therefore a change in the Chern-Simons number changes
the baryon number. However the
Chern-Simons number is not a meaningful physical quantity since
it is gauge
dependent and only changes in the Chern-Simons number
will be gauge invariant.

After the phase transition the $W_3^\mu$ and $Y^\mu$ mix to form the
massive $Z^\mu$ and the massless electromagnetic $A^\mu$ vector bosons.

In terms of these new vector fields Eq.\intg takes the form,
$$
\eqalign{
\quad \quad \Delta B = {N_f \over{32 \pi ^2}} \int d^4 x
[& \quad g^2  \vec E_W ^{\bar a}  \cdotp \vec B_W ^{\bar a} +
\alpha ^2 \cos 2\theta_w \vec E_Z \cdotp \vec B_Z +  \cr
{\alpha ^2 \over 2} & \sin 2 \theta _w (
    \vec E_A \cdotp \vec B_Z + \vec E_Z \cdotp \vec B_A )+
               i.t.] }
\eqn\mixf
$$
where $\alpha = \sqrt{g^2 + g'^2}$ and
the electric and magnetic fields are defined respectively,
$$
\eqalign{
{E_{\bar a}}^i _{W} = W_{\bar a} ^{0i}, \quad &
{B_{\bar a}}^i _{W} = {1 \over 2} \epsilon ^{i}_{jk} W^{jk}_{\bar a} \cr
E^i _{A} = A ^{0i}, \quad &
B^i _{A} = {1 \over 2} \epsilon ^{i}_{jk} A^{jk} \cr
E^i _{Z} = Z ^{0i}, \quad &
B^i _{Z} = {1 \over 2} \epsilon ^{i}_{jk} Z^{jk} }
\eqn\relt
$$
and $\bar a =1,2$. The last term of Eq.\mixf represents interaction
terms of the electric and magnetic components of $\vec Z$ and $\vec A$
with $\vec W _{\bar a}$ and
self-interactions of $\vec W _{\bar a}$.
{}From Eq.\mixf we see that there is no coupling
between the electric and magnetic components of $\vec A$.

Once the $W^1$, $W^2$ and $Z$ bosons become
massive they are exponentially suppressed unless there is some
topological obstructions in their configurations. We will assume that
after the phase transition we have $Z$-strings. At the core of the strings
we have false vacuum and the $Z$ bosons will be massless.
Equation\mixf can be reduced to
$$
\Delta B = {N_f \over{32 \pi ^2}} \int d^4 x \left [
\alpha ^2 \cos 2\theta_w \vec E_Z \cdotp \vec B_Z +
{\alpha ^2 \over 2} \sin 2 \theta _w
  ( \vec E_A \cdotp \vec B_Z  + \vec E_Z \cdotp \vec B_A )
\right] .
\eqn\ieee
$$
{}From Eq.\ieee we can infer a neat physical interpretation for the change in
the baryon number. The first term in the r.h.s. of Eq.\ieee ~
is the change in the helicity of the string network\refmark\gf.
The second term can be easily understood if we realize that
the background magnetic fields transform to electric fields in the
frame of the moving strings. Therefore this term is the interaction
of the magnetic $Z$ flux
along the core of the $Z$-strings with these electric fields.
The last term will have the inverse interpretation of the previous one.
The transformation of $\vec B_Z$ from the string to the background frame
will give us an $\vec E_Z$ field that will interact with the background
magnetic field.

To quantify the effect of the first term of Eq.\ieee would  require
to know, using a computer simulation,
the percentage of loops that are linked in the string network. We shall
not do this calculation\refmark\vf . In what follows we shall
assume that the contribution from this term is not greater than the
contributions coming from the other two terms.

Now that we have established a new mechanism to introduce
anomalous currents that will
violate baryon number, we need a process that biases these currents
towards the production of a net baryon number.
We are going to consider the two-Higgs model as an example. We shall show
that the classical equations of motion for the strings are modified by
the quantum corrections coming from the effective potential.
For collapsing strings, this corrections will favor the production of
baryons over antibaryons.

The relevant term in the effective action that biases the baryon number
is given by
the coupling of the relative phase $\theta$ of the two Higgs doublets
with the $W^a$ gauge bosons through the triangle diagram\refmark\tz.
For slowly varying $\theta$ this term can be expanded in a power series
in the momenta\refmark\dt .
We also do the same diagram but replacing the $W^a$ by the $Y$ gauge
fields. When we include both diagrams we find that the effective
action contains the term,
$$
\Delta S = \int d^4 x \kappa \theta {m_t^2 \over T^2}
                \left( g^2 W_a^{\mu \nu} \tilde{W} ^a _{\mu \nu}
                  - g'^2 Y_{\mu \nu} Y^{\mu \nu} \right)
\eqn\assi
$$
where $\kappa = {14 \over 3 \pi ^2 \zeta (3)}$ and the expansion is up
to quadratic order in $m_t /T$.
Using the standard definitions of $Z^\mu$, $A^\mu$ vector bosons
and assuming that the $W^1$ and $W^2$ are exponentially suppressed we obtain
$$
\Delta S \simeq {N_f \alpha ^2\over 64 \pi ^2} \kappa \sin 2\theta _w
        {m_t^2 \over T^2}
\int d^4 x \theta \left (\vec E_A \cdotp \vec B_Z +\vec E_Z \cdotp \vec B_A
\right ) .
\eqn\assp
$$
This term in the action violates $C$ and $P$ but
conserves $CP$. However the evolution of $\theta$ will not conserve
$CP$ because of explicit breaking of $CP$ in the two-Higgs potential.
This term will modify the classical equations of motion of the
string network.
As the string network is formed Eq.\assp ~ will bias the evolution
of the interactions of the magnetic flux of the strings with the
electromagnetic background. This will change the rate of baryon number
produced. However in order to favor the production of a net baryon number
we need that the angle $\theta$ changes in time in a definite direction,
so that the transition rate will be more favorable towards the production
of baryons over antibaryons.
The directionality of $\theta$
is given by the fact that the string network is collapsing. In particular
we have seen that we will have strings bounded by monopole-antimonopole
pairs and small loops that will be collapsing.
At the moving ends of the strings, $\theta$
and the Higgs fields that describe the strings will evolve in a definite way
in time.

Another way to understand this result is using the approach of Ref.\tz ~ .
We can consider that $\theta$ is
homogeneous in space and integrate by parts the first term in the
r.h.s. of Eq.\assi , we obtain
$$
\Delta S \simeq \kappa ' {m_t^2 \over T^2}
\int dt {d \theta \over dt} (N_{CS} - n_{CS}) ,
\eqn\assj
$$
where $\kappa ' = {448 \over 3 \zeta (3)}$.
This corresponds to a potential that favors a change in the Chern-Simons
number in a definite direction and because of \cher ~ we will obtain a net
change in the baryon number.

Now the only thing that is left to do is to show that
Sakharov's third condition is also satisfied.
We notice that the $M-\bar M$ pairs
and string loops
will collapse at relativistic speeds, this
will give baryon production with processes that are out of thermal equilibrium.

Finally we are going to calculate a rough estimate of the amount of baryons
produced.
Unfortunately, we do not understand the formation of electroweak strings
at the electroweak phase transition. There are some papers that study
the statistical mechanics of strings and monopoles connected by
strings\refmark\mt .
However they do not consider string interactions or break-up of strings
into monopole antimonopole pairs, so that their results are not applicable
to our problem. Also it seems that the Kibble mechanism will not apply.
However topology has to play an important role in their formation since
these strings are embedded defects.
If the Higgs field winds non trivially around the vacuum
manifold a string has to be produced if it is metastable. At present there
is not a reliable method to estimate the density of strings or their length
distribution. In order to obtain some numerical estimates we shall make
the optimistic assumption that the density of strings is one per correlation
volume.

The magnetic $Z$-flux along the
core of the string is $\sim 2 g T^2$. We also know that
the strings will be moving in a background magnetic field.
Refs.\tv ~ and \eo ~ show that the energy density of the primordial magnetic
fields scale as radiation, so that $B \sim g T^2$.
The magnetic fields will be smooth in scales of the
order of the correlation length $\xi \sim T$.
At this temperature the scale of $\xi$ corresponds to
the inter-particle separation of the background plasma. As the correlation
length grows the electromagnetic fields will start to be affected by the
fact that the plasma is a very good conductor. So we would expect that
the magnetic fields will be frozen in the plasma.

The transformation of these magnetic fields under a
Lorentz transformation from the frame of the background plasma to the system
of the string moving with velocity $v_t$ with respect of the background is
of the order of $E_A \sim \gamma v_t B_A $. As a first approximation we
have neglected the electric fields assuming that the plasma is a very
good conductor. The rate of baryon number produced per unit time and
volume is
$$
\Gamma \sim {N_f \over 64 \pi^2 } \alpha ^2 \sin 2 \theta_w
E_A \cdotp B_Z \sim \alpha _W ^2 \gamma v_t T^4
\eqn\amnr
$$
where $\alpha _W = g^2 / 4\pi \sim \alpha^2 /4\pi$.
This rate is three orders of magnitude higher
than the rate of baryon number produced through thermal transitions across
the sphaleron barrier ($\Gamma _s \sim \alpha _W ^4 T^4$).
This equation has to be modified to include the effect of corrections that
come from the effective potential.
As we saw before this contribution will be significant
only at the end of the strings or in the collapsing loops.
This will give that the volume in which
the correction is effective is $\sim \delta ^3$ where $\delta$ is the width
of the string ($\delta \sim \lambda ^{-1/2} \eta ^{-1}$). The rate of baryon
generation per string is
$$
{d N_B \over dt} \sim \alpha_W^2 \gamma v_t g^2 T^4 \epsilon \delta ^3
\eqn\amns
$$
where $\epsilon$ is a dimensionless constant that gives the numerical
contribution of Eq.\assp to the rate of baryon generation.
To find the rate of increase in the baryon number density
we have to multiply the previous expression by the density of strings.
If we assume one string per correlation volume at the time
of the phase transition ($\xi ^3\sim \lambda ^{-3} \eta^{-3}$),
we obtain
$$
{d n_B \over dt} \sim \alpha_W^2 \gamma v_t g^2 \lambda^{1/2} \epsilon T^4 .
\eqn\amnp
$$
This has to be integrated by the time that lasts the string network
($t_d \sim d/v_c \sim \lambda^{-1} \eta^{-1}$) so that
$$
n_B \sim \alpha_W^2 \gamma v_t g^2 \lambda^{1/2} \epsilon T^3 .
\eqn\amnq
$$
The entropy density at the time of the phase transition is
$$
s = {\pi ^2 \over 45} \mu T^3 ,
\eqn\llll
$$
where $\mu$ is the number of spin states. We obtain that the ratio
between the baryon and entropy density is,
$$
{n_B \over s} = {45 \over \pi ^2 } {1 \over \mu} \alpha_W^2 \gamma v_t
   g^2 \lambda ^{1/2} \epsilon .
\eqn\llli
$$
For $\lambda \sim 1$ we need $v_t \epsilon \sim 10^{-4}$ in order to
obtain the observational value of $n_B /s \sim 10^{-8}$.

{\bf{Acknowledgements}}

I would like to thank R. Brandenberger, G. Field, T. Vachaspati and
A. Vilenkin for helpful discussions.
I also thank the Ministry of Education of Spain for financial support.

\vfill
\eject

\vfill
\eject
\refout
\end